# Study of x-ray emission enhancement via high contrast femtosecond laser interacting with solid foil

**L. M. Chen, M. Kando, S. V. Bulanov, J. Koga, K. Nakajima, T. Tajima**

*Advanced Photon Research Center, Kansai Photon Science Institute,*

*Japan Atomic Energy Agency, Kyoto 619-0215, Japan*

**M. H. Xu, X. H. Yuan, Y. T. Li, Q. L. Dong, J. Zhang**

*Institute of Physics and China Academy of Sciences, Beijing 100080, China*

We studied the hard x-ray emission and the Kα x-ray conversion efficiency ($\eta_K$) produced by 60 fs high contrast frequency doubled Ti: sapphire laser pulse focused on Cu foil target. Cu Kα photon emission obtained with second harmonic laser pulse is more intense than the case of fundamental laser pulse. The Cu $\eta_K$ shows strong dependence on laser nonlinearly skewed pulse shape and reaches the maximum value $4x10^{-4}$ with 100 fs negatively skewed pulse. It shows the electron spectrum shaping contribute to the increase of $\eta_K$. Particle-in-cell simulations demonstrates that the application of high contrast laser pulses will be an effective method to optimize the x-ray emission, via the enhanced "vacuum heating" mechanism.

The availability of intense femtosecond laser pulses [1] opens a new laser-solid interaction regime in which intense laser pulses deposits onto a solid faster than the hydrodynamic expansion of the target surface. Hot electrons generated via collective absorption mechanisms such as resonant absorption (RA) [2] or vacuum heating (VH) [3] penetrate into the solid target to produce hard x-rays via K-shell ionization and bremsstrahlung [4]. This kind of intense and ultrafast hard x-ray source has a number of interesting applications for medical imaging techniques [5] because of its small x-ray emission size, its compactness and the shortness of its pulse duration.

Control and optimization of the hard x-ray emission produced by high intensity laser-solid interaction requests an understanding of several mechanisms: the laser energy absorption, the hot electron generation and the x-ray conversion. Several groups have already reported x-ray emission experiments relying on sub-picosecond laser systems [6-12]. Previous works [6,7] used hundreds of femtosecond laser pulse produced by $CO_2$ or Nd laser systems. Plasma density gradient steepened by ponderomotive force and satisfy the optimal conditions for RA, which is the main heating mechanism at this regime. Recently, it has shown that the use of shorter laser pulse durations less than 100 fs involve new x-ray emission processes. Eder *et al.* reported observing a maximum in K$\alpha$ emission when the target was placed away from best focus [8] and qualitatively explained with the re-absorption of produced photons inside the target. Based on optimal scale length for RA, Reich *et al.* [9] theoretically presented a scaling law to estimate the optimal laser intensity and predicted a reduction of the hard x-ray yield if the laser intensity is higher. Zhidkov *et al.* [10] studied prepulse effects with a low contrast fundamental 42 fs laser due to presence of ASE and showed the presence of a large plasma gradient $L/\lambda=2.5$ at modest laser intensities laser. They observed a decrease of the laser energy absorption for shorter pulse duration with constant laser energy, which was also proved by Schnürer *et al.* in experiment [11], and they reported the critical influence of the plasma gradient for the hard x-ray emission via the resonant process. All theses publications proved that there is a limitation for hard x-ray enhancement with laser intensity based on RA when tens of fs, low contrast laser are used.

In this letter, we show a breakthrough for this limitation of hard x-ray enhancement in the case of high contrast relativistic fs laser pulse. Cu Kα photon exhibits a higher flux when we work with a high contrast laser pulse at 400 nm. Enhanced "Vacuum heating" (eVH) mechanism should be stimulated in our experimental condition and in contribute to the enhancement of x-ray emission. The yield of x-ray emission can be controlled and optimized via detuning compressor gratings positively and negatively. The maximum $\eta_K$ can reach $4x10^{-4}$ for 100 fs negatively skewed pulse irradiation.

The experiments are realized with the high intensity Ti: Sapphire laser system on Lab of Optical Physics, Institute of Physics of China Academy of Sciences. The laser delivers a maximum output energy of > 300 mJ after compression with a pulse duration of 60 fs. After compression, the prepulse from 8 ns before the main pulse is better than $1x10^5$ monitored using fast photodiode. The laser contrast for picosecond pedestal obtained using s high dynamic range third-order femtosecond auto-correlator (Sequoia) is $1x10^4$ (see **Fig. 1a**). A type I potassium dideuterium phosphate (KDP) frequency doubling crystal (1 mm thick) is used to get the 400 nm second harmonic pulse. The double-frequency conversion efficiency of the KDP crystal is about 35% at 200 GW/cm$^2$ intensity. The infrared is almost rejected by passing the beam over 4 dielectric coated mirrors. This increase a pulse contrast ratio compared to the picosecond pedestal > $10^8$ and > $10^{10}$ in the ns time window. Finally the *p*-polarized laser pulse is obliquely incident on the target at $45^0$ by an f/3.5 parabola mirror in a focal spot diameter of 10 μm (FWHM) with an average $1x10^{18}$ $W/cm^2$ intensity. By means of a λ/2-plate *p*- or *s*-polarized light could be used. Cu foil target with thickness 5 μm was used in experiments. The measurement of the x-ray spectrum and the determination of the $\eta_K$ are made with a single photon counting x-ray LCX-CCD camera [6] placed 1.5 m away from source as a dispersionless spectrometer. An electron spectrometer with permanent magnetic field B=1000 Gs is used to detect electron spectrum [12]. Imaging plate is used as detector in spectrometer and also used for electron angular distribution measurement. The x-ray emission size is measured by the knife-edge imaging technique [6]. The FWHM of fitted Gaussian function shows source size ~10 μm in 400 nm laser irradiation, implying no evident plasma expansion in this case.

Pulse duration is increased by detuning compressor grating but then the uncompensated linear or B-integral phase reduces pulse contrast. We were able to vary the laser pulse duration from 60 fs up to 2 ps by changing the distance between gratings at constant laser energy. Distance increasing (decreaseing) will cause incomplete compensation of accumulated phase nonlinearities results in negatively (positively) chirp pulses having a gentle (steep) rise time [13], as shown in **Fig. 1b**. For a sample pulse, the integrated rising edge energy, normalized by the case of 60 fs, are 0.98 and 2.14 for laser with pulse duration 100 fs positively skewed and 100 fs negatively skewed respectively. It shows that positively skewed 100 fs pulses are provided with almost the same rising edge as the case of 60 fs laser pulse, whereas the negatively skewed one has double energy in rising edge. This is a key factor for following experimental phenomena. The development of preplasma for high contrast laser was calculated using hydrodynamic code

HYADES. A double Gaussian fit to the temporal pulse shape was used and an estimate of prevailing plasma scale length ($L/\lambda$) before the arrival of the high intensity pulse was obtained. It is 0.1, 0.05 and 0.04 correspond to 100 fs negatively skewed, 60 fs and 100 fs positively skewed pulse respectively. The effect of this scale length difference was investigated as following experiments and PIC simulations. To correctly interpret the data, it is crucial to establish the proper compressor zero which corresponds to shortest pulse duration. Group velocity dispersion in windows and mirrors affects the pulse duration at the detector as well as auto-correlator itself. This results in improperly determination of grating setting for minimum pulse width. The KDP crystal is used experimentally for this purpose. The energy conversion efficiency for second harmonic pulse is very sensitive to laser pulse duration and undesirable frequency chirp of the fundamental pulses at constant pulse energy. Frequency chirp of the fundamental laser pulses reduces the KDP energy conversion efficiency. As inset of **figure 2(b)** shown, KDP energy conversion efficiency is the maximum at the compressor grating position which we defined as compressor zero in experiment. This demonstrates our determination of grating zero is correct.

**Figure 2(a)** represents shows the spectra measured on a Cu target with an x-ray CCD camera with following parameters: 60 fs, 100 mJ, $1x10^{18}$ $W/cm^2$ at 800 nm (solid line) and 400 nm (dotted line) wavelength in p-polarized laser. We observe that the Kα yield at 400 nm is higher than at 800 nm by a factor of 2. The Cu $\eta_K$ in 2π steradian reaches ~$1x10^{-4}$ at this intensity. It should be noted, however, that the x-ray spectrum is not distributed in Maxwellian and an evident energy cutoff ($E \sim 20\ keV$) exist in inset of Figure 2(a), which also predicted in Ref. [10]. In order to optimize Cu $\eta_K$, we introduce slight long pulse duration with nonlinearly skewed pulse shape. **Figure 2(b)** represents Cu $\eta_K$ as a function of laser pulse width at negatively skewed (solid circle) and positively skewed (solid square). It shows Cu $\eta_K$ with negatively skewed 100 fs pulse width reach a maximum as $4x10^{-4}$ and almost 5 times greater than the case of positively skewed pulse. Inset shows KDP conversion efficiency has no dependence as a function of pulse duration with different chirps. However, for S-polarized laser incidence, Cu $\eta_K$ (dotted line) is 3 folds lower and do not show evident pulse chirp dependence. It simply reduces as increasing the laser pulse duration.

A higher $\eta_K$ for a 400 nm laser pulse corresponds to a higher laser energy absorption by hot electrons. According to Freshel's equations prediction and Price's experimental results [14], inverse bremsstrahlung (IB) decreases for an increasing intensity, whereas in our results the x-ray emission increases as a function of the laser intensity. Therefore, it implies that an additional absorption mechanism is stimulated. If we consider that RA is the main additional mechanism to generate hot electrons [2], a 800 nm laser pulse should be more effective than a 400 nm one because the later presents a much weak pedestal which induces a smaller L (<0.1λ) that is far away from optimal scale length for RA[15]. However, our measurement doesn't agree with this assumption. Simulation shows VH dominates RA for steep density gradient [3, 16]. VH means that a P-polarized light pulse is obliquely incident on an atomically abrupt metal surface in order to be strongly absorbed by pulling electrons into vacuum during an optical cycle, then returning to the surface with approximately the quiver velocity [3]. In our laser condition with intensity of $1x10^{18}\ W/cm^2$, the plasma scale length is 0.05λ according to our hydro-calculation and Ref [17] in similar conditions. It agrees with the necessary condition to stimulate VH: $X_{osc} \geq L$. It also satisfies the optimal condition for VH: $V_{osc}/c \geq 3.1(L/\lambda)^2$ [16] in which $V_{osc}=eE/m_e\omega$ is the electron quiver velocity in the laser field that is governed by the quiver energy:

$E_q=mc^2[(1+2U_p/mc^2)^{1/2}-1]$, where $U_p(eV)=9.3x10^{-14}I\lambda^2$ is the ponderomotive potential. The most important evidence is the cutoff energy we detected, i.e. 20 keV, which rationally fit for the scaling law of VH: $E_q=15\ keV$. The x-ray emission size we measured confirmed our electron energy measurement. The x-ray emission size generated by 400 nm laser is 15 ± 5 μm. This value is much smaller than in the case of 800 nm laser pulse ( > 80 μm) [12]. Therefore, we conclude VH is stimulated and may be the main absorption mechanism in our experiment.

Hot electrons generated by intense laser field are responsible for producing characteristic and bremsstrahlung x-ray radiation as they interact with a solid target. Therefore, it is important to determine the energy spectrum and angular distributions of fast electrons, which are generated by P- and S-polarized laser fields. As **figure 3** shown, fast electron emission is concentrated in target normal and specular reflection direction in case of S-polarized laser incident. For P-polarized laser irradiation, electron emission is much stronger, by a factor 4-5, and show broad emission in regime of concerned, excepting a peak along target direction that is presented, in Ref. [17], as surface fast electrons when similar high contrast laser incident. Fast electron spectrum shows a peaked structure and it is suitable for Cu Kα photon generation, considering Cu K-shell ionization cross section. The spectrum cannot be fitted by Maxwellian distribution for a temperature because the laser pulse is too short to cause plasma satisfied to local thermal equilibrium [12]. It should be noted that there is an individual second peak with energy ~ 110 keV in case of P-polarized laser irradiation, which means another group of quasi monoenergetic electron heated by other mechanism and contribute enhancement of fast electron and Kα photon generation in this case. Considering enhanced "vacuum heating" mechanism [18] as a candidate, which stimulate surface plasma wave resonantly excited by laser and accelerated electrons perpendicular to target surface, it will accelerate quasi monoenergetic electron, when ponderomotively heated electron phase match for charge separation potential, to maximum energy $E_{max}$ ~ $mc^2(\gamma_{osc}-1)$ ~ 120 keV in our experimental condition, that is almost same as experimental data.

Simulations using a 1D fully electromagnetic LPIC++ code have been performed, where an electromagnetic wave is launched obliquely from the left-hand side onto an over dense plasma located on the right-hand side. We used the following simulation parameters: $n_e/n_c$=20, $T_e$=100ev, $T_e/T_i$=3~5 and mass ratio $m_i/Zm_e$=1836. The initial scale length is $L/\lambda$=0.05. A square-sine profile for the incident laser pulse is used. Typically 150x2,680 electrons and ions were used for 2,680 cells. We consider the initial situation in which the ions are mobile. **Figure 4(a)** shows the temporal dependence of the amplitudes of electric field in the normal direction close to the solid surface. It clearly shows a group of electrons is pulled out into vacuum at each optical cycle and return to target surface. The electric field at the target surface, which polarity is changed periodically, reflected this "pull-push" procedure in each laser half period [3, 16]. Electrons will absorb laser energy continuously in this way. It demonstrates the VH is greatly stimulated.

**Figure 4(b)** shows integrated electron energy absorbed depend on plasma density gradient (0~1) at $\Delta t$=30. In the range we concerned, there are 3 absorption peaks that locate at scale length $L/\lambda$=0.1, 0.25 and 0.6 respectively. According to optimal scale length, the second peak corresponds to resonant absorption [2]. The first one ($L/\lambda$=0.1) is result from Vacuum heating according to Ref. [16]. The curve shows the stimulation of VH is very critical depend as a function of L. This is a key character for VH that contribute to x-ray emission enhancement and its chirp dependence at high contrast laser irradiation.Experiment and simulation [16, 19] proved that, with a slight surface expansion of the scale length, the optical field would pull more electrons into vacuum and thus be more strongly absorbed, as long as L doesn't significantly exceed $X_{osc}$. In our experiments with constant laser energy, pulse pedestal will be expand if we tune compressor gratings negatively and match the optimal plasma gradient ($L/\lambda \sim 0.1$) for VH when pulse width is 100 fs, resulting in enhancement of $\eta_K$ to a maximum. Positively skewed pulse with same pulse width exist shorter pedestal with smaller (~40%) integrated energy at pulse pedestal according to our measurement, see Figure 1(b). in this case, the stimulation of VH is not so strong compare to the case of negatively skewed. That is the explanation for x-ray emission dependence with different laser pulse shape. We need to mention this hard x-ray laser pulse shape dependence have not observed in low contrast, 800 nm laser irradiations. On the other hand, foil target show advantages for K$\alpha$ photon enhancement because a strong electric field always exists on the rear side of target, see **Figure 4(c)**. Fast electrons that thrill through the foil will been drawn back by this field and return to foil again to produce more K$\alpha$ photons.

In conclusion, Cu $\eta_K$ produced by a high contrast laser pulse at 400 nm with intensity I = $1x10^{18}$ $W/cm^2$ reaches $4x10^{-4}$, thanks for restructure hot electron spectrum shape. Tuneable control of hard x-ray emission is succeed via control pulse duration and nonlinearly skewed pulse shape. It implies an effective method for hard x-ray enhancement in fs-plasma regime: resonant absorption maybe non-effective with femtosecond laser-dense plasma interactions [12], whereas high contrast laser is more efficient for hard x-ray generation via enhanced "vacuum heating".

This work is joint supported by the Trilateral project, KAKENHI project in KPSI JAEA and NSFC (Grant No. 10474134) in China.

## Captions

Figure 1: (a) Temporal pulse distribution for low-contrast 800 nm pulse (dashed line) and the high-contrast 400 nm pulse (solid line). The 800 nm pulse distribution is probed by SHG cross-correlator. The 400 nm pulse distribution is estimated from $I_{(400)} \sim I^2_{(800)}$. (b) Temporal pulse shape for high contrast laser with 60 fs (solid line), 100 fs positively skewed (dotted line) and 100 fs negatively skewed (dashed line) respectively.

Figure 2: (a) Cu hard x-ray spectra measured with a CCD camera produced by 800 nm (solid line) and 400 nm (dotted line) laser pulse at $1x10^{18}$ $W/cm^2$. (b) Cu Kα X-ray conversion efficiency as a function of laser pulse width at negatively skewed (solid circle) and positively skewed (solid square) in P-polarized laser, and also negatively skewed (open circle) and positively skewed (open square) in S-polarized laser. Inset shows KDP conversion efficiency dependence as a function of pulse duration with different nonlinear pulse shapes.

Figure 3: Hot electron detection in case of P- and S-polarized laser irradiation. (a) raw imaging on IP to show electrons. (b) Angular distribution of hot electron observed in case of P-polarized laser (solid line) and S-polarized laser (dashed line) interaction. (c) Hot electron spectrum observed in case of P-polarized laser (solid line) and S-polarized laser (dashed line) interaction.

Figure 4: (a) (color) temporal dependence of the amplitude of electrical field $E_x$. The critical surface start from $x=10.3$. (b) Integrated electron energy dependent as a function of density gradient ($L/\lambda$). P-polarized laser with 30 optical cycles and $1x10^{18}$ $W/cm^2$ incident on target at $45^0$. (c) The amplitude of the oscillating longitudinal electric field at $t$=30 optical cycles of the laser field. The strong electric field exists at the foil target rear-side. The electric field here is in normalized units of $m\omega_0 c/e$.

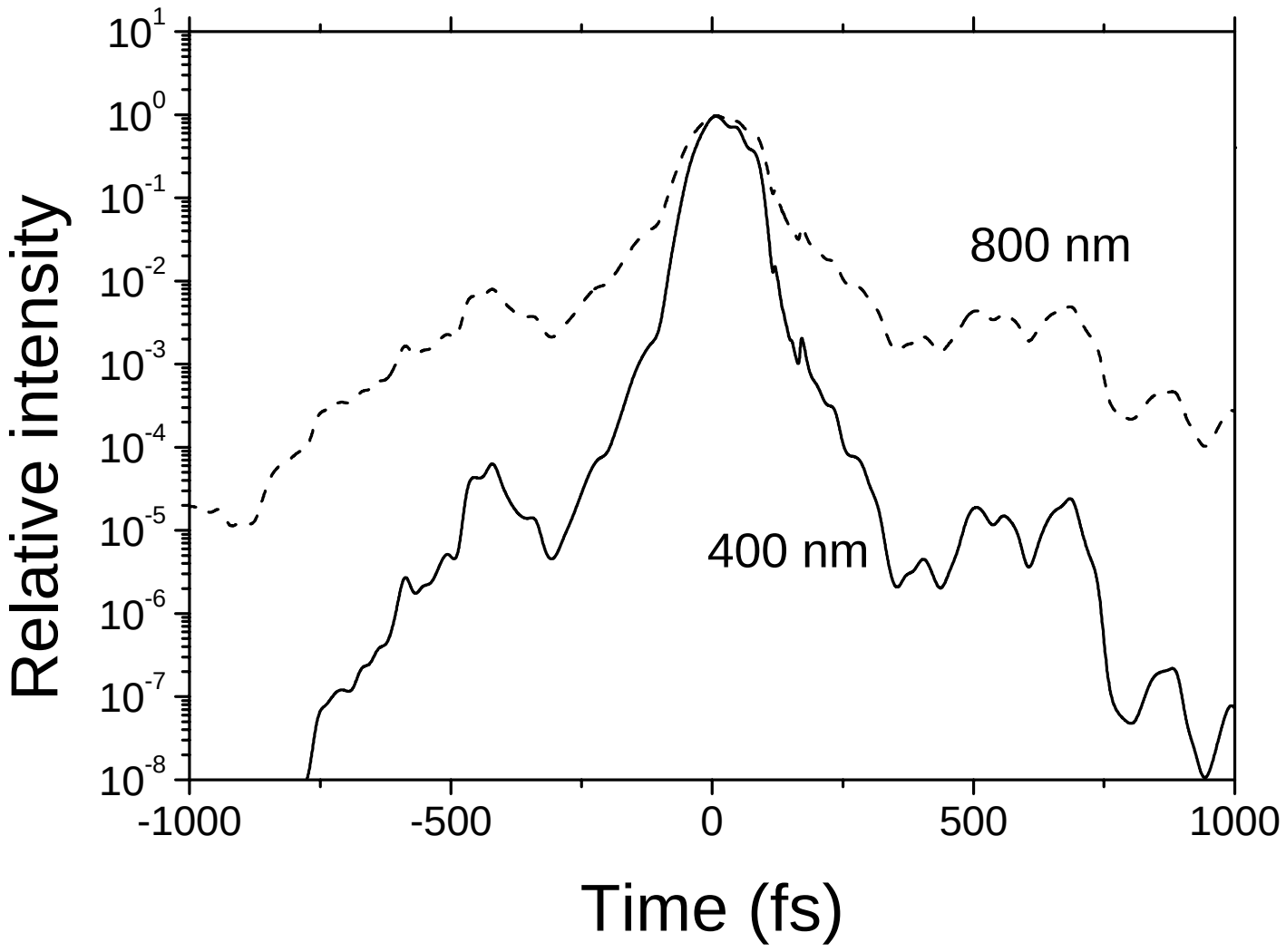

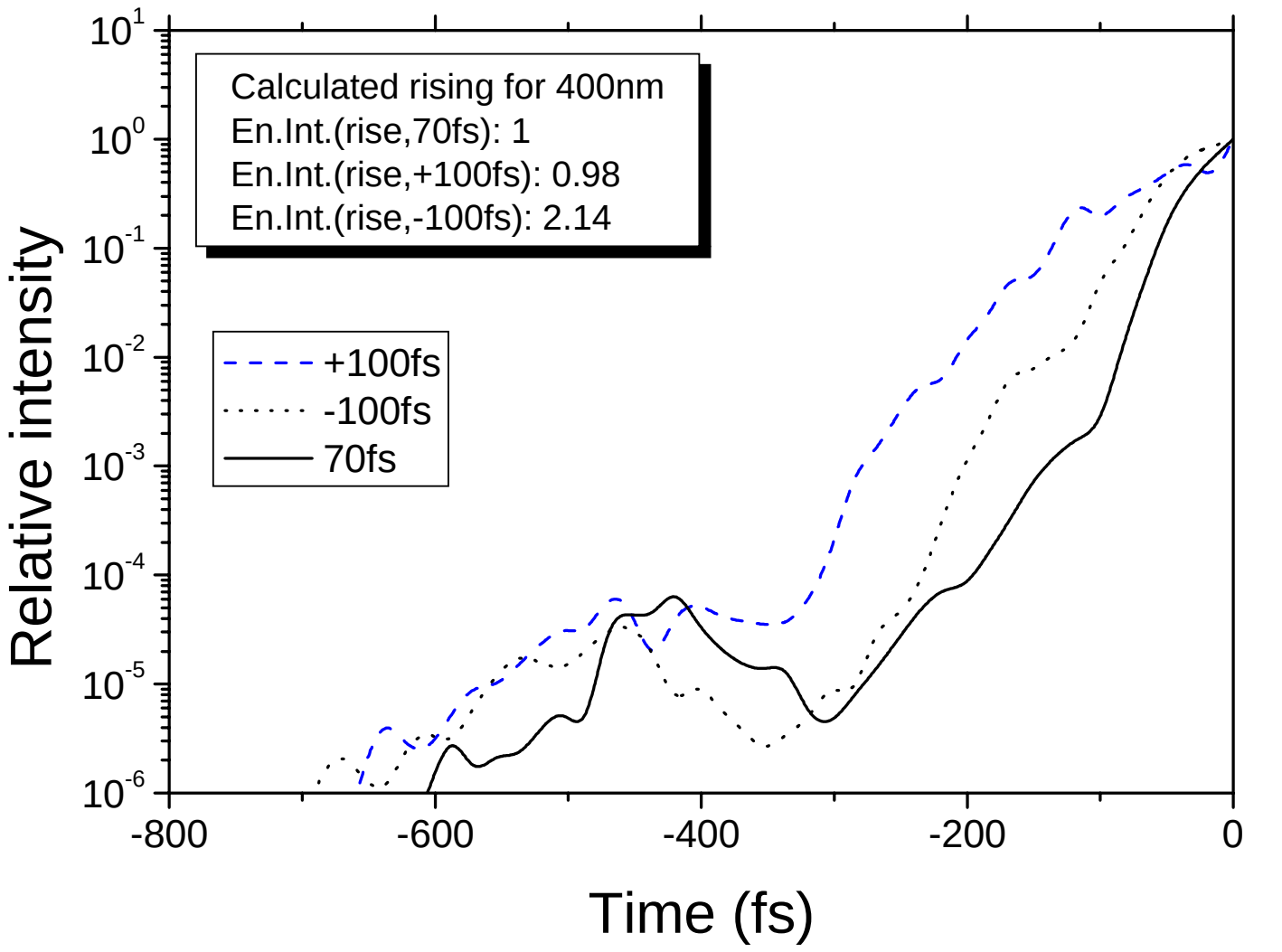

Figure 1: L. M. Chen et al.

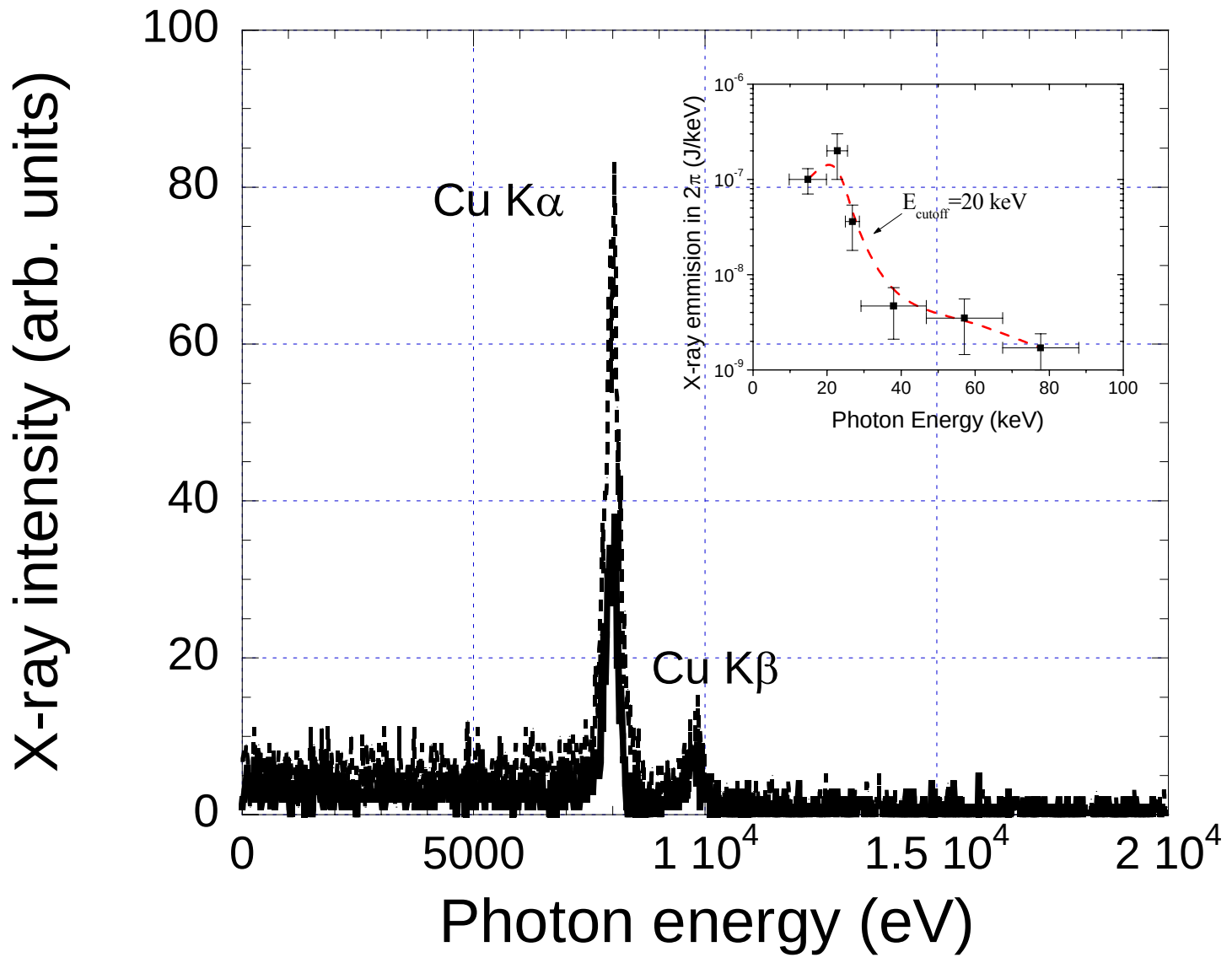

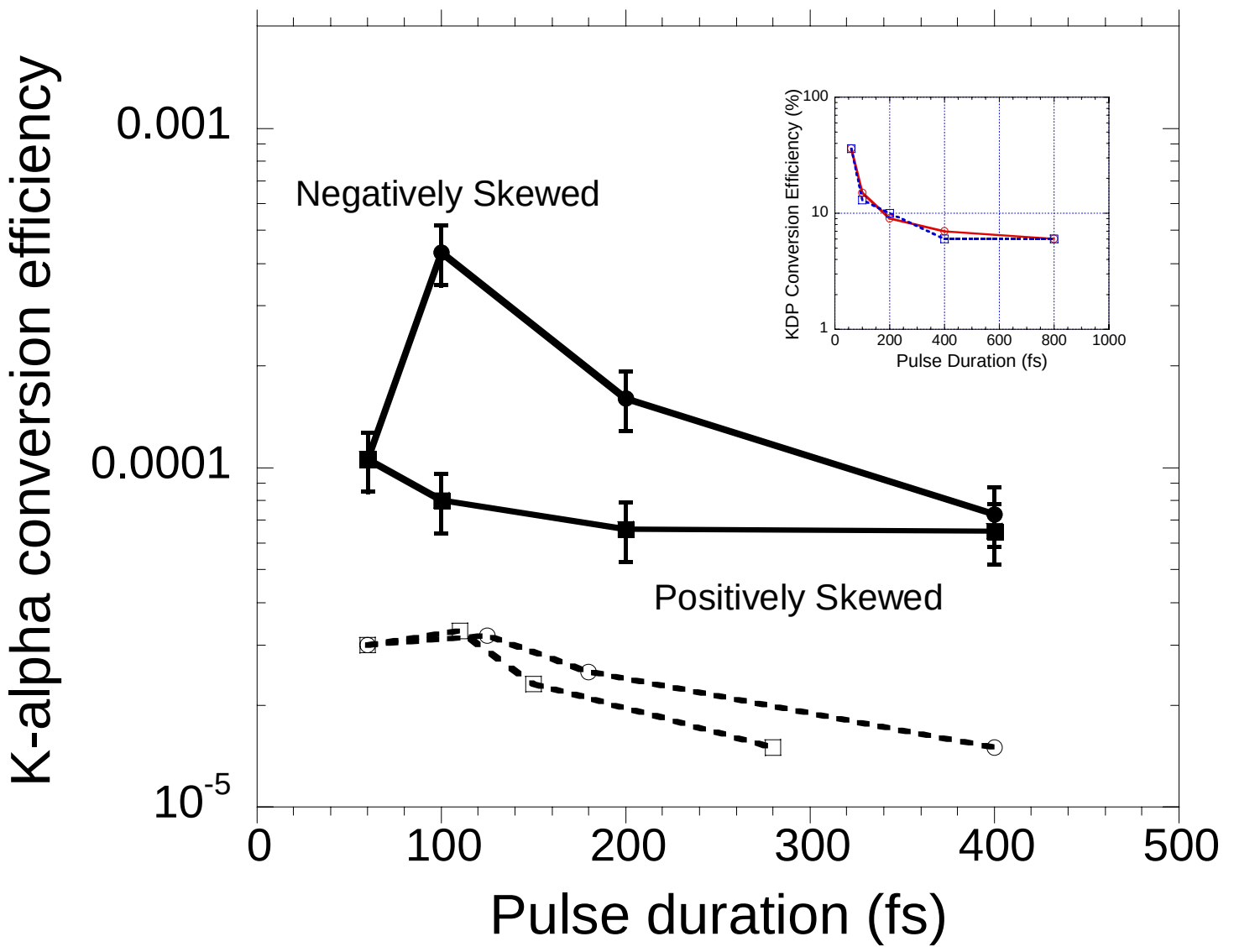

Figure 2: L. M. Chen et al.,

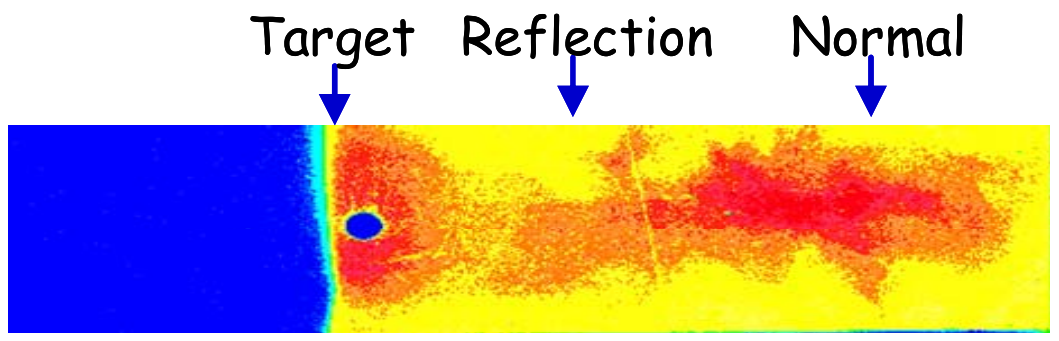

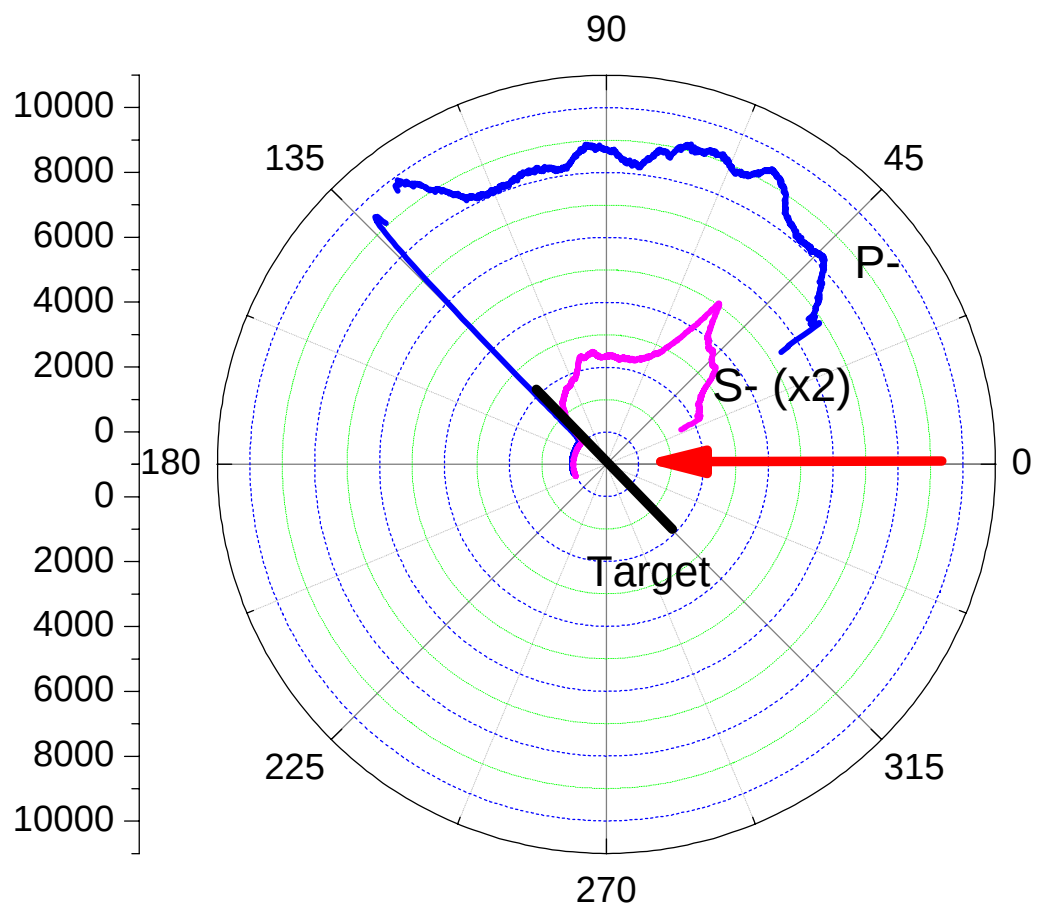

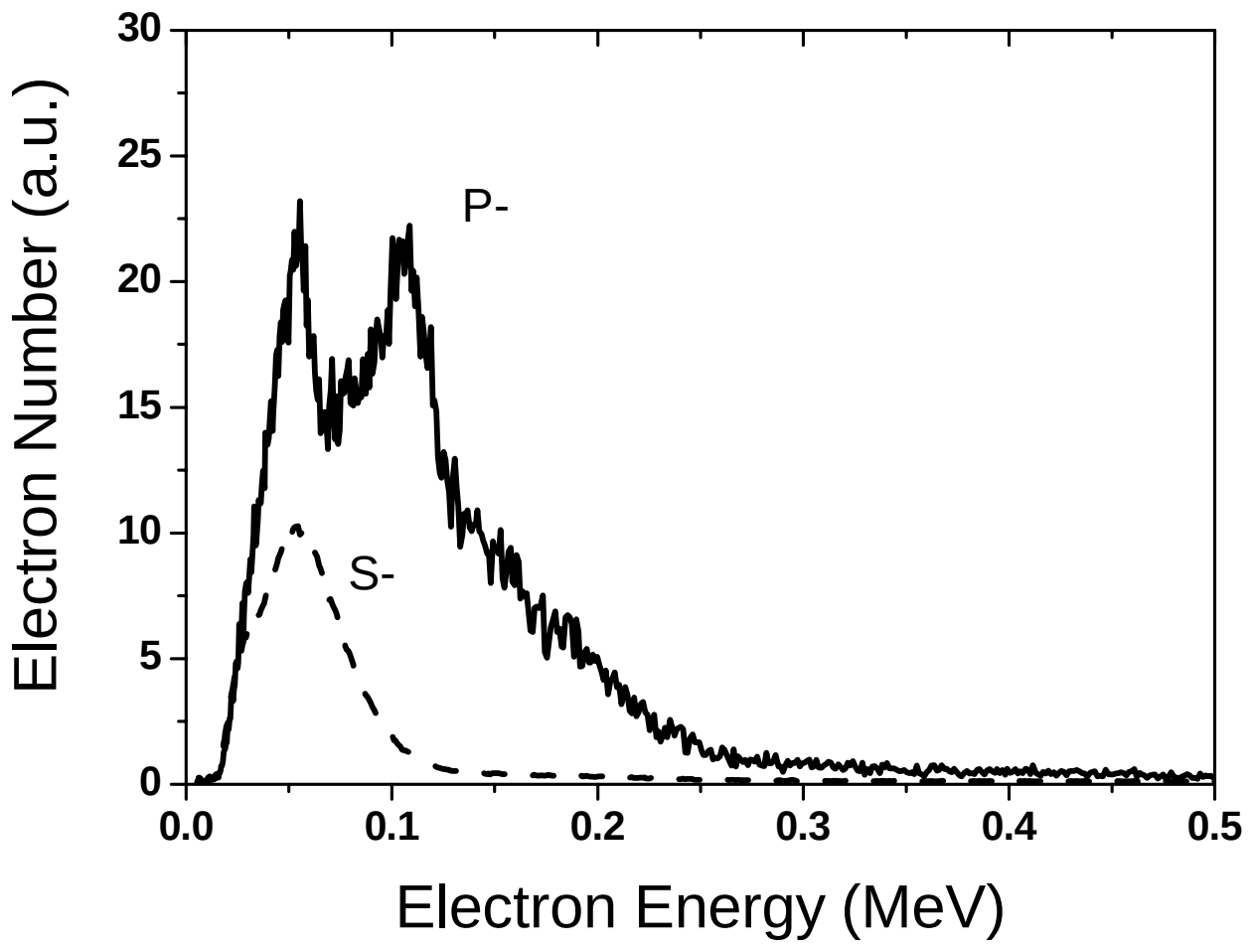

Figure 3: L. M. Chen, et al.,

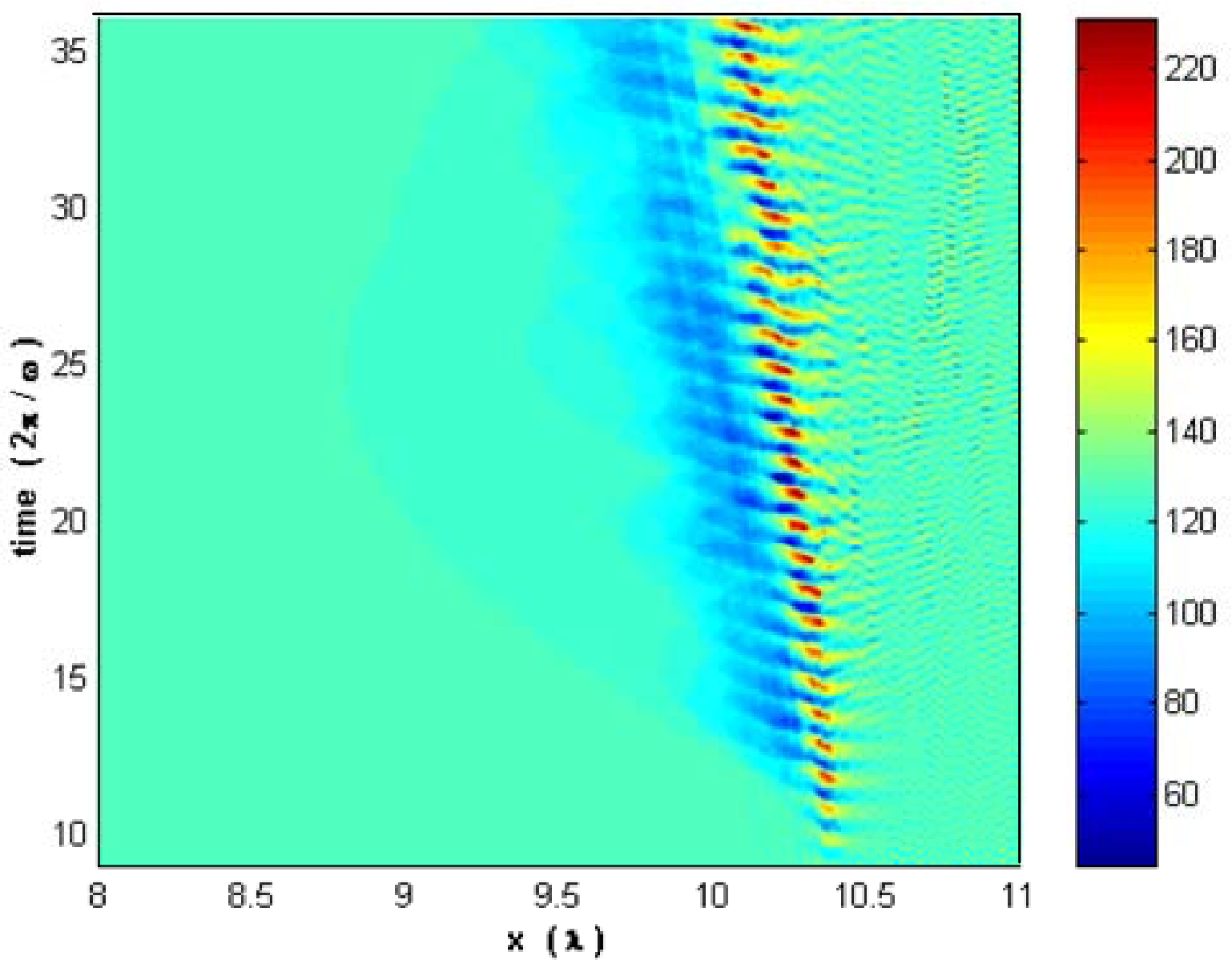

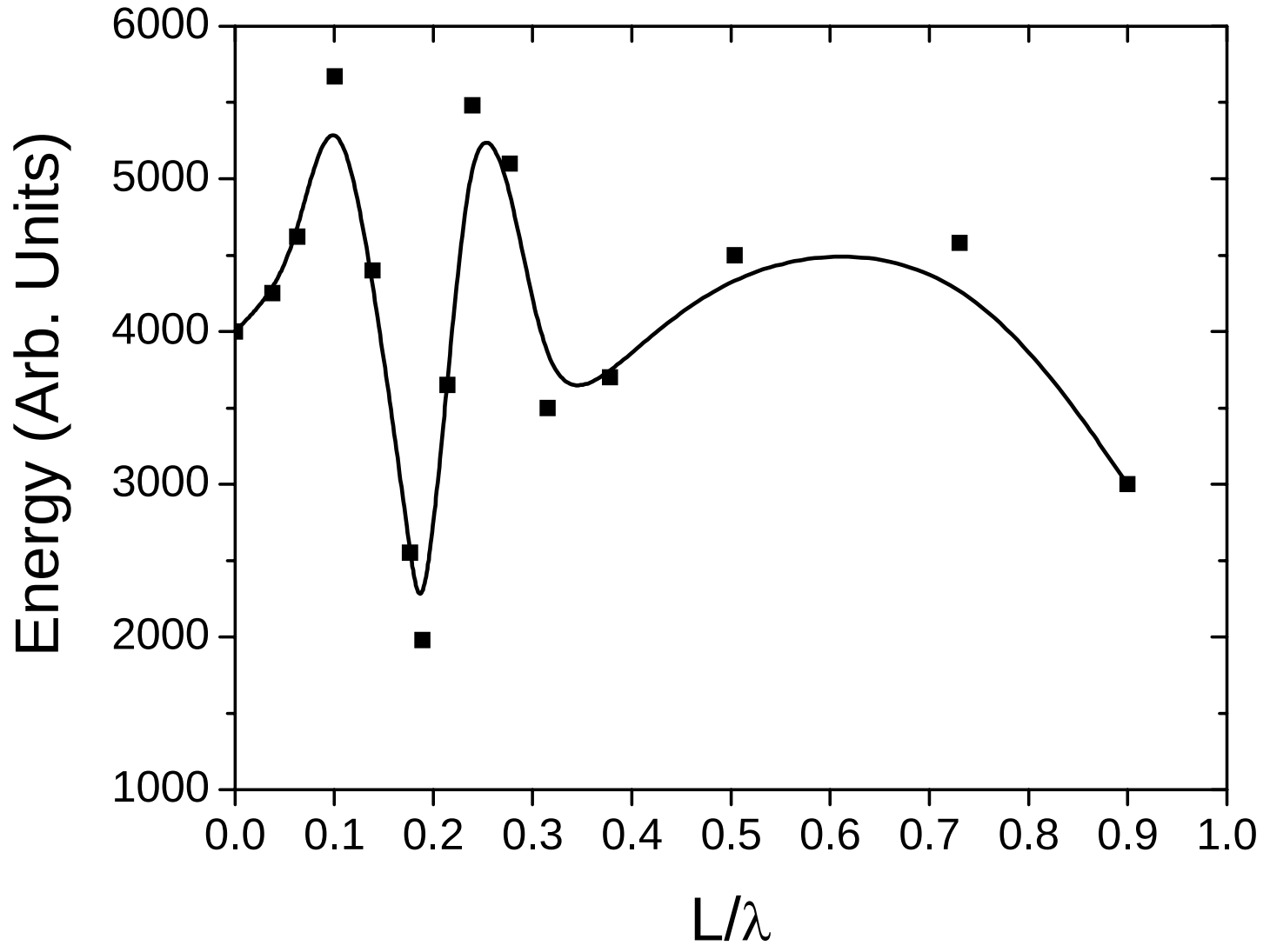

Figure 4: L. M. Chen et al.,

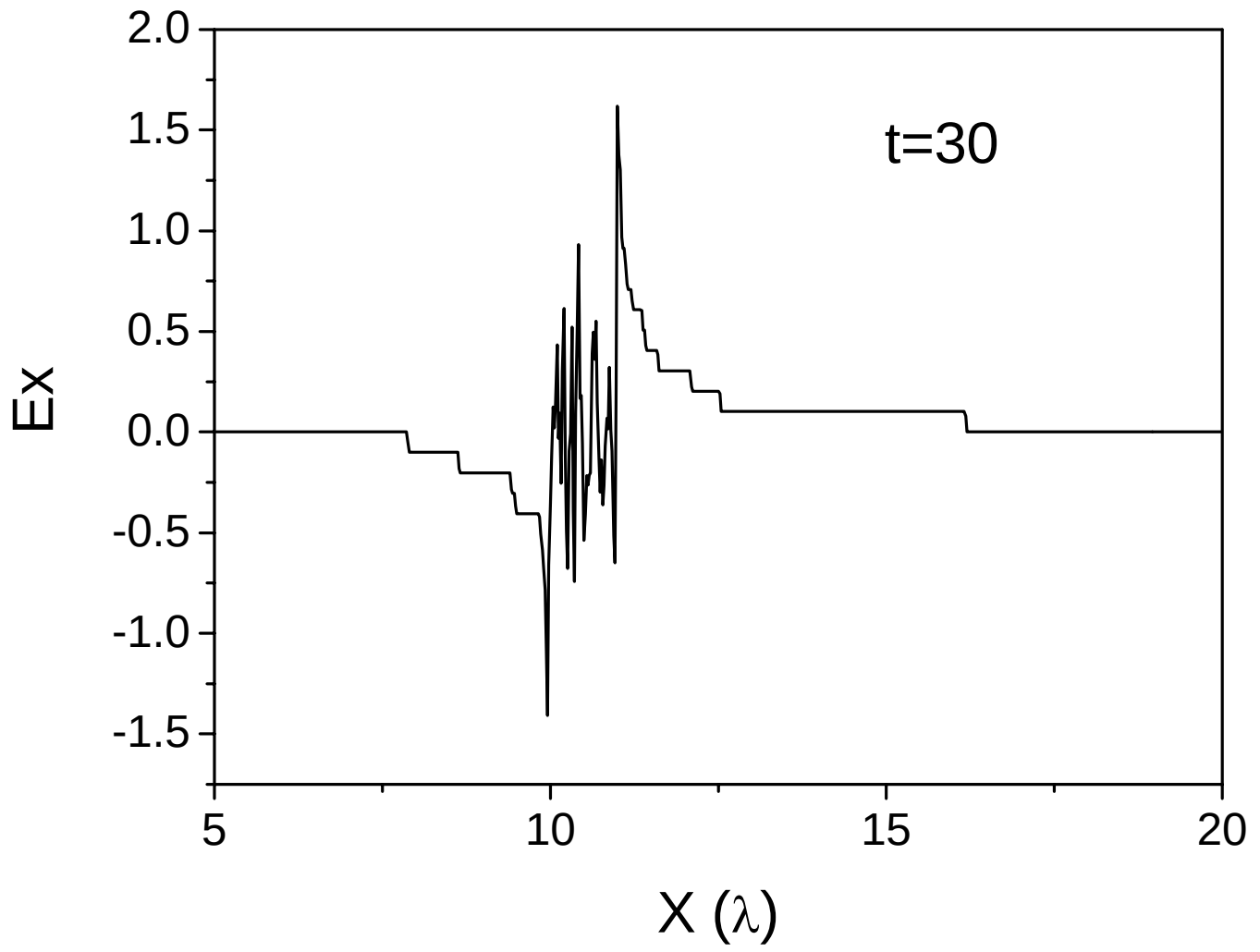

Figure 4: L. M. Chen et al.